\pdfoutput=1

\PassOptionsToPackage{hyphens}{url}

\documentclass[sigconf]{acmart}

\usepackage{booktabs} 

\usepackage{titlesec}

\begin{document}
\title{The Value of Software Architecture Recovery for Maintenance}

\author{Daniel Link}
\affiliation{
  \institution{University of Southern California}
  \city{Los Angeles}
  \country{U.S.A.}
}
\email{dlink@usc.edu}

\author{Pooyan Behnamghader}
\affiliation{
  \institution{University of Southern California}
  \city{Los Angeles}
  \country{U.S.A.}
}
\email{pbehnamg@usc.edu}

\author{Ramin Moazeni}
\affiliation{
  \institution{Santa Clara University}
  \city{Santa Clara}
  \country{U.S.A.}}
\email{rmoazzeni@scu.edu}

\author{Barry Boehm}
\affiliation{
  \institution{University of Southern California}
  \city{Los Angeles}
  \country{U.S.A.}
}
\email{boehm@usc.edu}

\renewcommand{\shortauthors}{D. Link et al.}

\begin{abstract}
In order to maintain a system, it is beneficial to know its software architecture. In the common case that this architecture is unavailable, architecture recovery provides a way to recover an architectural view of the system. Many different methods and tools exist to provide such a view. While there have been taxonomies of different recovery methods and surveys of their results along with measurements of how these results conform to expert's opinions on the systems, there has not been a survey that goes beyond a simple automatic comparison. Instead, this paper seeks to answer questions about the viability of individual methods in given situations, the quality of their results and whether these results can be used to indicate and measure the quality and quantity of architectural changes.
For our case study, we look at the results of recoveries of versions of Android, Apache Hadoop and Apache Chukwa obtained by running PKG, ACDC and ARC.
\end{abstract} 



\copyrightyear{2019} 
\acmYear{2019} 
\setcopyright{acmcopyright}
\acmConference[ISEC'19]{12th Innovations in Software Engineering Conference (formerly known as India Software Engineering Conference)}{February 14--16, 2019}{Pune, India}
\acmBooktitle{12th Innovations in Software Engineering Conference (formerly known as India Software Engineering Conference) (ISEC'19), February 14--16, 2019, Pune, India}
\acmPrice{15.00}
\acmDOI{10.1145/3299771.3299787}
\acmISBN{978-1-4503-6215-3/19/02}

\maketitle
\section{Introduction}

The promise of software architecture recovery is that it yields results that are not only accurate, but also help the stakeholders of a system to evaluate the system and to estimate the impact possible changes of the system would have on its architecture.
This holds true without any regard to the specific paradigm any recovery method's view is based on.
Proving the accuracy of a recovery result is relatively easy. Verifying its utility is more complicated, but it can certainly be shown that some accurate results have limited utility.

One reason for the interest in architecture recovery is to gain information on whether the architecture fits a certain desired style or exhibits anti-patterns such as architectural smells \cite{Garcia2009}.

Once such determinations have been made, there may follow an interest of the stakeholders to rearrange the architecture to fit the desired style or to fix its issues. This can happen from one version of the system to the next, or over a sequence of more than two versions. Architecture recovery will be helpful in this process if it can show the stakeholders whether they are on the right track.

For any metric or view of a system produced by an algorithm to be a true aid in system maintenance, it will have to take into account that system maintenance is commonly performed (1) incrementally \cite{Mohagheghi2005}, (2) by teams \cite{MockusAudrisandTFieldingRoyandDHerbsleb2002}, and (3) distributed and on different parts of the system \cite{Herbsleb2003}.

\begin{figure}[b]
    \centering
    \includegraphics[width=0.39\textwidth]{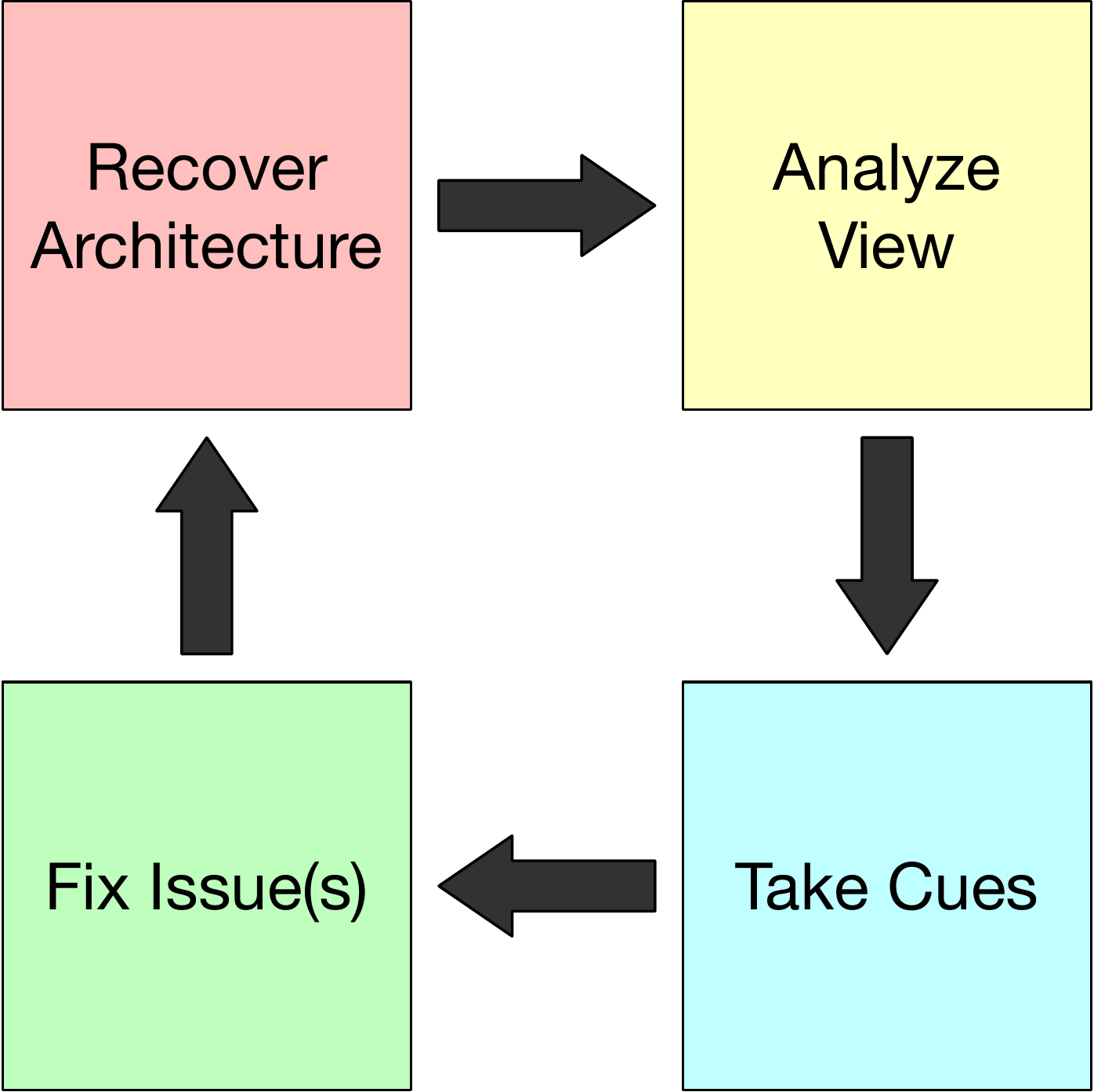}
    \caption{The Virtuous Cycle of Architecture Recovery}
    \label{fig:virtuouscycle}
\end{figure}

For the recovery to aid in such a maintenance process, it has to necessarily fulfill several criteria, some of them very basic. The following enumeration lists them, along with reasons why we believe they have to be fulfilled.
\begin{itemize}
\item Architecture representation: The result needs to show a structure, and some kind of reasoning about the structure must either be visible or possible to be deduced. Otherwise it does not provide an architecture (refer to Section \ref{foundation_architecture} for our definition of Software Architecture).
\item Feasibility: It has to be possible to recover the architecture within a finite and predictable amount of time. If this is not the case, architectural recovery becomes an open-ended task and unreliable as an aid. When the recovery is not open-ended, other relevant questions are how long recoveries take in absolute time as well of how a method scales to large systems. 
\item Clarity: The output needs to explain how and why the system's entities were grouped. For this, the result needs to be either self-explanatory or provide the user with some clear information they can refer to for further inquiries. This is required so the user can take an educated guess of how their possible modifications could affect the architecture. Clarity also helps in communicating results within a team.
\item Proportionality: The severity of a modification between two given versions should be adequately reflected in the recovery results. This means that changes to the system should be reflected in the new recovered architecture in proportion to the extent that they affect the attributes of the system that form the basis of the paradigm. The alternative is that maintenance becomes an unpredictable process.
\item Determinism: Recovery results should be reached in a deterministic manner. A lack of determinism makes it not only impossible to reproduce the results, but also to tell if any differences in the architectural view between two versions stem from changes made by the maintainers or instabilities of the recovery algorithm. Non-determinism can also put the accuracy of a recovery method in question when it leads to significantly different results on each new run.
\item Continuity: If two different architectural views represent the beginning and the end of a planned process of one or more modifications to the system's architecture, and it is possible to reach an intermediate state between the two, then the architectural view of the system in the intermediate state should reflect this by being recognizable as being ``between'' the two extreme points. The reason for this is that maintainers will not want to know if their changes are setting the system on the correct path to the desired goal.
\item Isolation: It needs to be possible to change only one element in isolation, and have this reflected in the recovered architecture. (Note that what this element is depends on the paradigm on each recovery method.) This is important for enabling distributed work on the system without incurring undesirable crosstalk between changes.
\item Accuracy: The resulting architectural view needs to be intrinsically true to the system itself or some architectural ``ground truth'' view of the system that can be arrived at by other means than running the recovery method.
\end{itemize}

To the extent that we can show that some or all of these conditions are not satisfied, a given recovery method will be of limited use for the purposes of incremental development.\\
We have therefore for the first time studied how several available recovery methods that have been used for studies such as \cite{Garcia2013},\cite{Lutellier2015}, \cite{Le2015} and \cite{Behnamghader2016} measure up against the criteria listed above. We have not directly measured their accuracy but instead addressed the utility of ``ground truth'' architectures.

The structure of the rest of this paper is as follows: Section \ref{foundation} elaborates on the foundation of our research. Section \ref{related} discusses work related to ours. Section \ref{approach} introduces our research approach. Section \ref{results} presents our results. Section \ref{threats} covers threats to validity. Sections \ref{conclusion} and \ref{future} conclude this paper and name our plans for the continuation of this research. \section{Foundation}\label{foundation}

\subsection{Software architecture and its recovery}\label{foundation_architecture}
Before we settle on a definition of architecture, we need to consider what architecture recovery is and what it can do.
Architecture recovery is the process of recovering a system's architecture from its implementation artifacts, such as its source code.
In many cases, such as the ones under review here, the result is a set of clusters that contain artifacts of the system, typically its source files.
This result aims to represent a view of the architecture under a paradigm espoused by the respective architecture recovery method. While one widely accepted definition of \textit{software architecture} as ``the set of principal design decisions about a system''\cite{Taylor2010}, there is a potential mismatch between the architecture of the system under this definition and the view that can be obtained through architecture recovery: Principal architecture decisions may not have been implemented and are therefore out of reach of architecture recovery. Conversely, the system may reflect inadvertent decisions that have never been made explicit, but are nonetheless present. Finally, some decisions may not be embodied in the artifacts or attributes thereof that a given recovery method considers or only emerge when the system is used in a certain way.

For this reason, we think that a definition that is more in line with the capabilities of programmatic architecture recovery is that of ``fundamental concepts or properties of a system in its environment embodied in its elements, relationships, and in the principles of its design and evolution'' \cite{ISO/IEC/IEEE}. This is the case because the definition includes what is embodied in a system's elements, such as source code, that the recovery can be based on. However, both may be of use to analyzers of the system.

\subsection{Concerns in Software Engineering}
The concept of a \textit{concern} is used throughout software engineering literature, frequently without any definition. It is generally agreed upon that a separation of concerns is desirable. 
Out of the several meanings the word ``concern'' can have in the English language, the ones most suitable for software engineering are the noun defined as ``matter for consideration'', 
and the verb defined as ``to be of interest or importance to'' \cite{Merriam-Webster}. Defined in this way, a concern can be viewed as something that one or more human beings want to exist or happen, and which can be expressed in natural language. Applied to a software system, a concern is something the system needs to do or to have, such as a functional or non-functional property. This is in line with the definition of it as ``a software system's role, responsibility, concept, or purpose'' used in \cite{Garcia2011}.

\subsection{Topic Modeling and Latent Dirichlet Allocation} \label{foundation_LDA}
The aim of probabilistic topic models is to discover the thematic structure in large archives of documents \cite{Blei2012}. One advantage is that the algorithms for probabilistic topic modeling do not need any annotations or labeling of these documents because the topics emerge from their analysis. Probabilistic topic modeling analyzes a body of text to find groups of words which are likely to occur together. Such groups of words are then presented as topics. 

Latent Dirichlet Allocation (LDA) \cite{Blei2003} is a statistical model based on the intuition that documents exhibit multiple topics with a different proportion, where each word is drawn from one of the topics, and where the selected topic is chosen from the per-document distribution over topics.

\subsection{Stop Words}
Topic modeling requires lists of stop words for its input. Stop words are words in a language that are very common, but do not have any meaning by themselves, such as ``the'', ``and'' and many others. Such words are like noise in that they carry low information content, cause low retrieval rates and have no predictive value. They are commonly grouped into either the general or domain specific category \cite{Makrehchi2008}. As the term ``domain-specific'' implies, they differ from domain to domain. Words like ``data'', "compute" and ``code'' can be stop words in software engineering (because nearly every piece of software deals with data, computations and software in some way) and would be useless in most software related concerns, but keywords in the automotive field, where they hold informational value.

\subsection{ACDC}
The ACDC (Algorithm for Comprehension-Driven Clustering) algorithm \cite{tzerpos2000acdc} uses structural relationships specified as patterns to create an algorithm for recovering components and configurations that bounds the size of the cluster (the number of software entities in the cluster), and provides a name for the cluster based on the names of files in the cluster. ACDC's view is oriented toward components that are based on structural patterns (e.g., a component consisting of entities that together form a particular subgraph). It bears mentioning that ACDC was not introduced as an architecture recovery method, but as a clustering algorithm with the goals of facilitating program comprehension and coming up with system decompositions that are close to those of experts for given systems \cite{tzerpos2000acdc}. The official implementation of ACDC is written in the Java language \cite{Shtern2010}. It has been re-implemented for inclusion in the ARCADE workbench (see Section \ref{arcade}), omitting options for (1) reordering its patterns, (2) excluding some of them, (3) limiting the size of clusters in subgraphs, (4) using a hierarchical decomposition instead of a flat one, (5) including only top level clusters in the flat decomposition as well as (6) graphical output of the decomposition. (In order to evaluate recovery results in a way that is comparable with other research, such as the one in, we used a different version of ACDC that has been adapted for and integrated into ARCADE (see Section \ref{arcade}. In the further course of this paper, this is the version of ACDC we are evaluating in places such as Section \ref{commonalities}.)

\subsection{ARC}\label{foundation_arc}
ARC (Architectural Recovery using Concerns) uses topic modeling to find concerns and combines them with the structural information to automatically identify components and connectors \cite{Garcia2011}. For this, it leverages LDA (see Section \ref{foundation_LDA}) to find concerns and compute the similarity between them \cite{Garcia2011}. LDA can detect concerns in individual code entities and compute similarities between them. For this, the software system is represented as a set of documents called a corpus. Individual documents within it are ``bags of words''. Each document can contain different topics, which stand for concerns. In the output, topics are represented by the words that are most likely to appear in them, in descending order. It is also determined how relevant a topic is to each document in the corpus. Documents (representing implementation entities) are clustered using structural information (dependencies in the case of ARC) and concerns (the topics from the topic model) as features. The number of topics and clusters is set via parameters.
ARC's view aims to produce components that are semantically coherent due to sharing similar system-level concerns (e.g., a component whose main concern is handling of distributed jobs).
ARC has been implemented in Java\cite{Garcia2018}.

\subsection{PKG}
PKG \cite{Le2015} is very simple in that it only recovers the package-level structure view of a system's implementation. It produces an objective but not architecturally satisfying view in that it stays at the surface instead of trying to assist its user to determine why the system is built the way it is. PKG has been implemented in Python \cite{ComputerScienceDepartment2017}.

\subsection{Commonalities of PKG, ACDC and ARC}\label{commonalities}
In addition to the results obtained through their own different recovery algorithms, all three recovery methods also rely on dependencies among source code entities and report them in their output in order to enable further processing (e.g., smell detection). These dependencies are detected using the Classycle library \cite{Elmer2014}. In the case of Java source code, this is done from the compiled class files.

\subsection{Measures of Architectural Similarity}\label{sim_measures}
The following measures compare two architectures in terms of how similarly they cluster the entities that make up a system, such as source files. For all three measures, the results will be a value between 0\% (no similarity) and 100\% (equality).
It is important to keep in mind that these comparisons are purely structural. Differences in the underlying reasons for these structures (e.g., concerns in the case of ARC or programming patterns for ACDC) are not measured.

\subsubsection{MojoFM}
MojoFM \cite{ZhihuaWen2004} is an effectiveness measure for software clustering algorithms. It assumes that both architectures contain the same entities and is therefore not suitable for studies in software evolution, but can still serve to compare the distance between two different views produced by different recovery methods or to determine the extent of non-determinism within the same method. MojoFM assigns values between 0 (no similarity) and 100 (identity), with higher values meaning greater similarity.
\subsubsection{a2a}
This new measure \cite{Behnamghader2016} has been created for evolutionary studies. It is based on a distance measure that determines the number of transformations necessary to get from one clustering to the other. Like MojoFM, a2a assigns values between 0 (no similarity) and 100 (identity), with higher values indicating greater similarity.

The minimum-transform-operation (mto) is the minimum number of operations needed to transform one architecture to another:
\begin{equation}
\label{formula:mto}
\begin{split}
mto(A_1,A_2) =  remC(A_1,A_2) + addC(A_1,A_2)~~~~~~~~~~~~~~~~~~~~~~~~~~~~~~\\
~~~~~~~~~+ remE(A_1,A_2) +  addE(A_1,A_2) + movE(A_1,A_2)
\end{split}
\end{equation}

The five operations used to transform architecture $A_1$ into $A_2$ 
comprise
additions ($\mathit{addE}$), removals ($\mathit{remE}$), and moves ($\mathit{movE}$) of implementation-level entities from one cluster (i.e., component) to another; as well as additions ($\mathit{addC}$) and removals ($\mathit{remC}$) of clusters themselves.

Note that each addition and removal of an implementation-level entity requires two operations: an entity is first added to the architecture  and only then moved to the appropriate cluster; conversely, an entity is first moved out of its current cluster and only then removed from the architecture.

The mto is normalized to calculate a2a:

\begin{equation}
\label{formula:a2a}
\mathit{a2a}(A_1,A_2) =(1-\frac{mto(A_1,A_2)}
{mto(A_{\emptyset},A_1) + mto(A_{\emptyset},A_2)}
) \times 100\%
\end{equation}
where $mto(A_{\emptyset},A_i)$ is the number of operations required to transform a ``null'' architecture $A_{\emptyset}$ into $A_i$.

\subsubsection{cvg} The cluster coverage measured by cvg \cite{Behnamghader2016} shows to which extent components that exist in one clustering exist in another.
Cluster coverage (cvg) indicates the extent to which  two architectures' clusters overlap. In other words, cvg allows engineers to determine the extent to which certain components existed in an earlier version of a system or were added in a later version: 

\begin{equation}
	\label{formula:cvg}
	\mathit{cvg(A_1,A_2)} =\frac
	{ \mathit{|simC(A_1,A_2)|} }
	{ |C_{A_1}| } \times 100\%
\end{equation}
where $|C_{A_1}|$ is the number of clusters in architecture $A_1$.

$\mathit{simC(A_1,A_2)}$ returns  the subset of $A_1$ clusters that have at least one ``similar'' cluster in $A_2$:
\begin{equation}
	\label{formula:simC} 
	simC(A_1,A_2)=\{ { c_i~\mid~ 
			c_i \in A_1,~\exists c_j \in A_2,~  
		\mathit{c2c(c_i,c_j)} > \mathit{th_{cvg}}} \}
\end{equation}
where 
$c2c$ \citep{Garcia2013}
measures the degree of overlap between the imple-mentation-level entities contained within two clusters.
More specifically, 
$\mathit{simC(A_1,A_2)}$ returns $A_1$'s clusters for which the $c2c$ value is above a threshold $\mathit{th_{cvg}}$ for one or more clusters from $A_2$.

\subsection{Code Smells and Architectural Smells}

Code smells are anti-patterns in programming that are associated with bad design or bad programming practices \cite{Emden2002}.
While code smells are not necessarily code issues or bugs and do not indicate their presence, it is likely that they will lead to increased maintenance effort.

An architectural smell is a commonly (although not always intentionally) used architectural decision that negatively impacts system quality \cite{Garcia2009}. Code smells are related to code entities, architectural smells are related to components. One does not automatically map to the other\cite{Le2018}.

\subsection{ARCADE} \label{arcade}
ARCADE is a collection of tools that offers (1) a collection of architecture recovery methods (including PKG, ACDC and ARC) (2) detection of architectural smells, (3) metrics of architectural change and decay and (4) correlations of implementation issues and architectural ones \cite{Garcia2014}. It is implemented in the form of two Eclipse \cite{TheEclipseFoundation2017} projects, in the Java and Python languages respectively. Different branches emphasize automatic processing of large numbers of systems and their versions or ease of use through a GUI.
PKG and ARC have been developed with and for ARCADE. ACDC has been adapted to ARCADE. All recovery methods expose some parameters directly (such as input and output directories in the file system) or require the user to modify their code (such as the number of topics in ARC or the order in which different versions of a system are evaluated for evolutionary studies) \cite{ComputerScienceDepartment2017}

The tools in it have been used in for several comparative and evolutionary studies \cite{Le2015} \cite{Behnamghader2016} \cite{Shahbazian2017}.
Since we are referring to parts of those studies in our evaluations of the recovery methods, we are using their implementations as found in ARCADE.

\subsection{Ground-Truth Architectures}
The need to evaluate the accuracy of architectural views produced by recovery algorithms has sparked a desire for ``ground-truth'' architectures that can serve as reference views 
that these views can be compared to. The accuracy of the view is then determined by the closeness of the view to that reference view. While such references views have been obtained without the systems' engineers \cite{Bowman1999} \cite{Mattmann2015a}, more recently there have been efforts to involve them in the process \cite{Garcia2013}.
It bears mentioning that in the latter case, the process has been aided by use of existing architecture recovery techniques. \section{Related Work}\label{related}

\begin{figure}[t]
\centering
\includegraphics[width=0.40\textwidth]{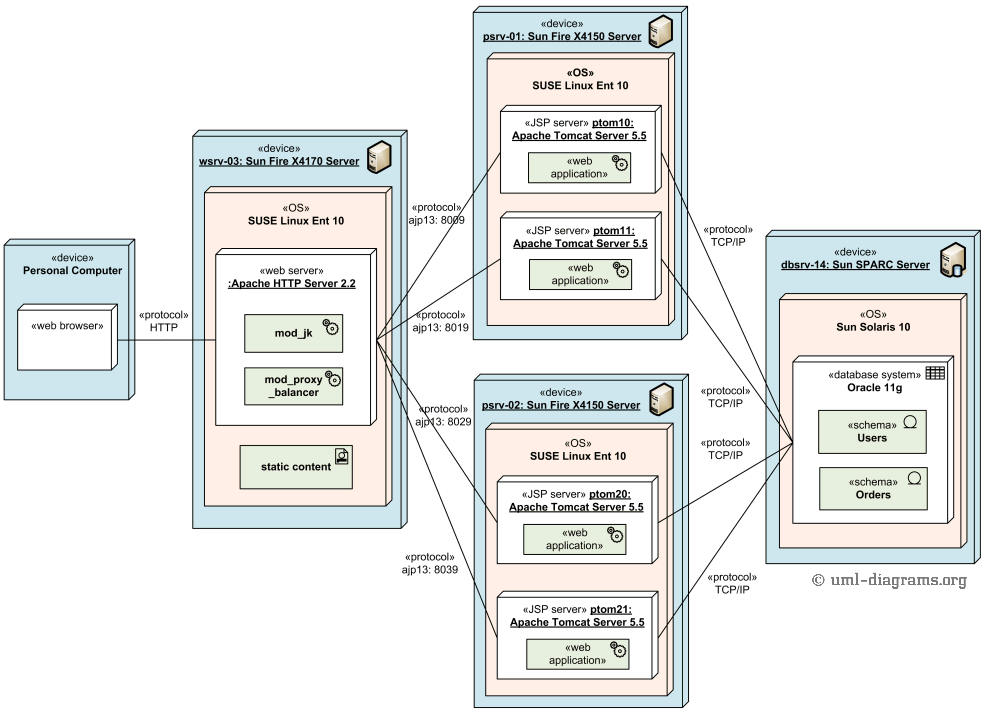}
\caption{UML Deployment Diagram}
\label{fig_deployment}
\end{figure}

Garcia et al. \cite{Garcia2013} have evaluated the results that six different recovery methods achieved for eight systems on how close they were to ``ground truth'' reference architectures available for each respective system. For this, they compared the clusterings produced by the recovery methods to the expert decompositions using MojoFM. However, notably, they experimented with letting each recovery method produce different amounts of clusters in order to arrive at the best possible MojoFM values the recovery methods could produce. In the case of ARC, they additionally started with setting the number of concerns to 10 and then raising it until a further increase did not improve the MojoFM values or they had reached a number of V/3 concerns, with V being the number of terms in a system. For another method that they knew to be non-deterministic, they also took the best result of three that it produced \cite{Garcia2013}.\\ 
In light of how several measures were taken in order to present all recovery methods in the best possible light and to pick and choose from their results, it is our conclusion that this kind of analysis of the ideal behavior of these recovery methods cannot be representative of how accurate any of these algorithms are when there is no ``ground truth'' to iteratively fit their results to. In this much more typical case, the user will not have an expert decomposition of the system, which is the reason to run architectural recovery in the first place.
Additionally, the way these ``ground-truth'' architectures are produced cause them to have four intrinsic limitations:
\begin{enumerate}
\item They rely on a comparison of clusterings of the same system. Conceivably, a recovery method could produce a conceptual view of a system which does not make use of clusters. (Consider a recovery method that bases its recovery exclusively on the semantics gained from parsing a system's documentation or the deployment of a system and would then come to a view similar to the the UML deployment diagram \cite{fakhroutdinov_2013} shown as an example in Figure \ref{fig_deployment}.)
\item The view they produce is necessarily affected by the paradigm of the existing recovery method the system expert uses as an aid. This can give an advantage to views generated with the same method if we consider that some recovery methods may use default values for the amount of clusters they generate (such as ARC, see section \ref{foundation_arc}).
\item The measure used to determine the distance between two architectural views limits how meaningful that comparison is. For example, while in concern-based recovery, the ground-truth architecture may assign one set of concerns to each cluster, none of the common similarity measures mentioned in Section \ref{sim_measures} take the semantics or names of concerns into account, but just clusters and their structures.
\item Human involvement introduces a measure of subjectivity to the process.
\end{enumerate}
Behnamgader et al. \cite{Le2015}\cite{Behnamghader2016} have used the recovery methods ACDC, ARC and PKG to study how the architecture of software systems changes between different kinds of versions (major, minor, patch). As a justification for using PKG, ACDC and ARC as implemented in ARCADE, they referred to the analysis mentioned above \cite{Garcia2013}, but did not take the iterative approach to running recovery methods espoused there, using pre-set parameters instead. Both ACDC and ARC were found to produce non-deterministic results. The non-determinism of ACDC, which was caused by the use of non-deterministic data structures in its ARCADE adaptation was addressed by pre-sorting the package names supplied to its orphan adoption code.
The non-determinism of the LDA algorithm employed by ARC as implemented in the MALLET toolkit was addressed by assigning a constant value as the seed to the random number generator in MALLET. Additionally, they observed that the impact of any changes to the system could not be predicted. They fixed this by generating a shared topic model for all versions of a system which are considered in an evolutionary study. 

\begin{figure}[t]
    \centering
    \includegraphics[width=0.50\textwidth]{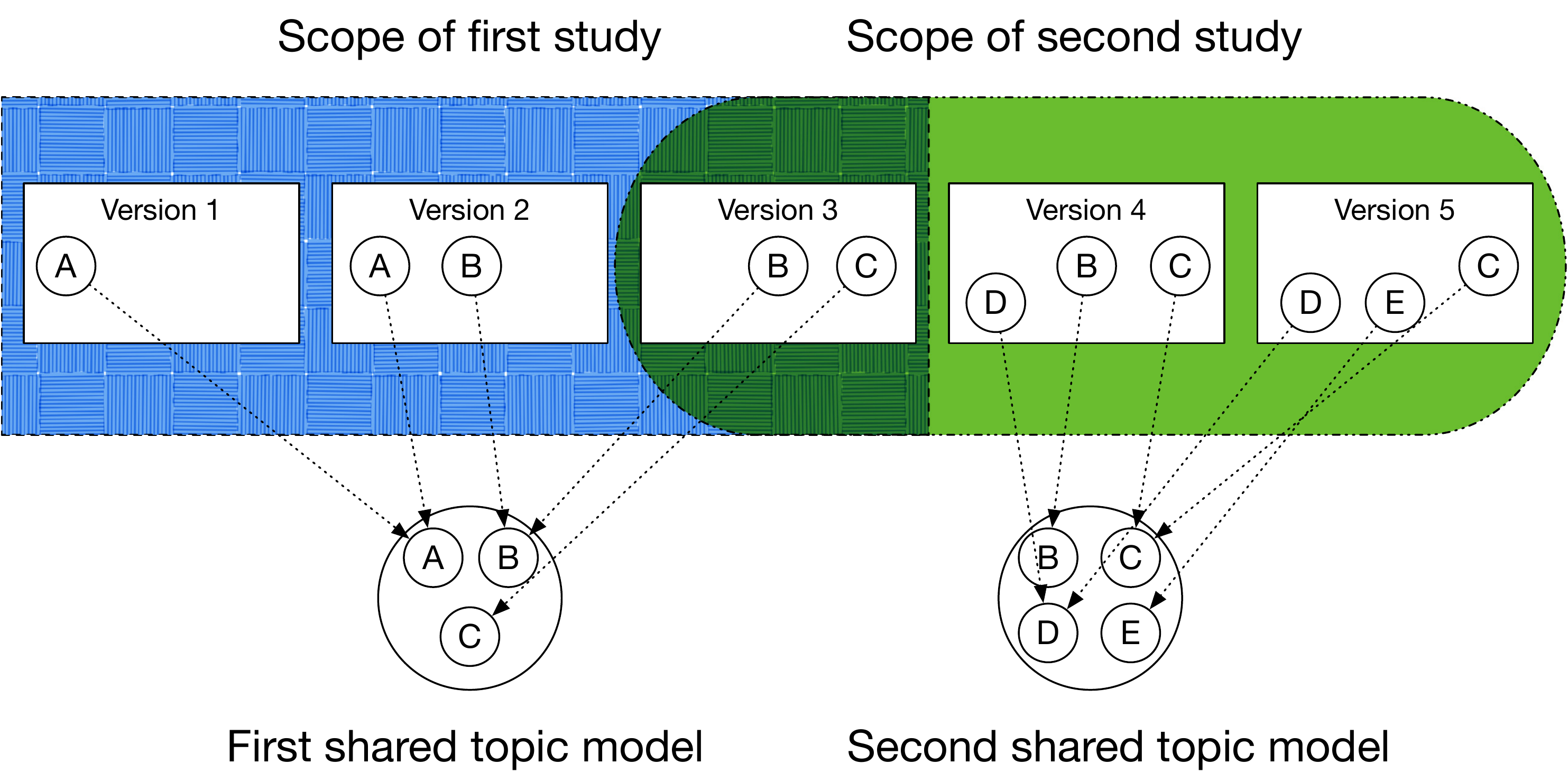}
    \caption{Shared Topic Models and Scope}
    \label{fig:sharedmodel}
\end{figure}

This shared topic model used for ARC has several deficiencies which are illustrated in Figure \ref{fig:sharedmodel}: Consider five consecutive versions 1-5 of a software system. Over time, the system has addressed concerns A through E as shown. Two evolutionary studies of the system are conducted. The scope of the first one covers versions 1-3, the second one versions 3-5. The issues of the shared topic model become apparent when we try to determine the correct architecture for version 3. Which concerns should be considered for its recovery? Looking at version 3 in isolation, we would consider concerns B and C. In the first study, we would apply a shared model made up of concerns A, B and C. In the second study, it would consist of concerns B, C, D and E. The former case would attest that version 3 addresses concern A when it clearly does not, while the latter would do the same with concerns D and E. It is also easy to see that for a system with n versions, 2\textsuperscript{n} different topic models could be built depending on which versions are included.
Lutellier et al. \cite{Lutellier2015} have studied six architecture recovery methods with a focus on their respective distances to ``ground truths'' and their processing times, but did not focus on our other concerns.

 \section{Approach}\label{approach}
\subsection{Research Questions}
For each selected recovery method, we asked the following questions:
\begin{itemize}
\item \textbf{RQ1:} Does the output represent an architecture?
\item \textbf{RQ2:} How feasible is it to recover a given system? 
\item \textbf{RQ3:} Does the output explain itself clearly?
\item \textbf{RQ4:} Are changes in the output proportional to those in the input?
\item \textbf{RQ5:} Is it deterministic?
\item \textbf{RQ6:} Are results continuous?
\item \textbf{RQ7:} Can change be isolated?
\end{itemize}

\subsection{Selection of Recovery Methods}
For practical reasons, we have limited ourselves to a sample of recovery methods which

\begin{itemize}
\item Have working implementations available
\item Have their source code available
\item Can be run on current commodity hardware and operating systems
\item Did not require us to purchase any software licenses
\item Are well-documented
\end{itemize}

Additionally, we favored recovery methods that had been evaluated against ground-truth architectures.

For this reason, out of the recovery methods mentioned in \cite{Garcia2013}, we chose PKG, ACDC and ARC. We have not been able to run Bunch \cite{Mancoridis1999} and Limbo \cite{Andritsos2005}. While there is an apparently working implementation of Bunch available, it relies on another tool called chava to generate its input from a Java system's source files. We have been unable to locate that tool.

\subsection{Selection of Subject Systems}
Our criteria were for selecting systems were that a system be:

\begin{itemize}
\item Well-known or systems have had recovery methods run on them in published literature
\item Written in Java or C/C++ (so ACDC and ARC could work on them)
\item Having meaningful words in their source code (important for ACDC and ARC)
\item Able to be compiled without modifications
\item Open-source
\end{itemize}

For this reason, we have selected Google Android \cite{Google2018}, Apache Hadoop \cite{Apache2018} and Apache Chukwa \cite{TheApacheSoftwareFoundation2016}.

\subsection{Computing Environment and Parameters}
In order to not impose tight limitations on the recovery methods, they were run on an 8-core Xeon system with 64 GB of main memory.

All recovery methods were run without modifications of their source code and their default settings, unless otherwise indicated. Since a recovery with ARC requires setting the number of desired concerns as a parameter, 100 was chosen for all systems. \section{Results}\label{results}

\subsection{RQ1 (Architectural Output)}
For the architectural view that the output of a recovery method constitutes to be considered an architecture, it has to reveal a structure as well as a reason for the structure (compare to Section \ref{foundation_architecture}). 
When a recovery method clusters source code entities, it will always produce a structure. To what extent it represents an architecture then depends on the reasoning for the structure it provides. The richer that reasoning is, the better the case for the recovery method being architectural becomes.
\subsubsection{PKG}
The output of PKG clusters the system's code entities by package, reflecting the package structure of the system. No attempt whatsoever is made to reveal the reasoning behind creation of the packages and the placement of the code entities in them. Considering also that packages are incidental to any software system, we cannot consider the output of PKG to represent an architecture.
\subsubsection{ACDC}
ACDC clusters code entities based on subsystem patterns and tries to assign meaningful names to the clusters. While the programming patterns constitute a reason for the structure, they themselves are incidental to the programming process. With the exception of systems serving as examples for learning system building, systems are not designed to exhibit programming patterns, but rather these patterns occur necessarily with the system's functionality. To the extent that the names ACDC assigns its cluster reveal meaningful information about the system's design, they reveal the reasoning behind the structure. Therefore, ACDC will produce architectural output when they do. This puts it a step above PKG.
\subsubsection{ARC}
ARC clusters code entities based on topics derived from words found in them. The clusters provide a structure, and the topics provide reasoning behind the structure as long as they are meaningful. 
\subsection{RQ2 (Feasibility)}\label{feasibility}
How feasible it is to use a recovery method depends on how much time and space is needed for it to be run on different existing systems and come to a satisfyingly accurate architectural view. This can be observed in absolute terms, and can be predicted on its time and space complexities. Rather than analyzing each algorithm found in the recovery methods for their complexity, we will make some simpler observations: (1) If the recovery method crashes when applied to one or more software systems, then it cannot be applied to all systems. (2) If a method needs to be re-run an indeterminate amount of time, then its runtime cannot be predicted and is not guaranteed to have an upper bound. (3) If the recovery method is guaranteed to produce an accurate result after the first run, we can experiment with running it on several systems of different sizes.
\subsubsection{PKG}\label{feasibility_pkg}
PKG bases its view solely on the packages found in a system and their contents. This simply reflects the unambiguous information found in a Java system's class files. There is no room for possible differences between two runs in which source code entities and packages exist and which entities the packages consist of. Therefore no need to re-run it in order to come to better results, since no variations in its results can exist. Running it once on any system will suffice.
\subsubsection{ACDC}\label{ACDC_feasibility}
In order to optimize its accuracy, ACDC has to be re-run an unknown amount of times (see Section \ref{related}). Since the expert decompositions necessary to assess its accuracy were not available, we could not compare the relative accuracies of different results and ran it once for each of our selected systems. It bears mentioning that ACDC crashed repeatedly when we attempted to recover the current version of Android with it. Since we were able to recover smaller systems with it, we suspect that the considerable size of the Android source code was the culprit. We have been unable to establish a relation between available hardware resources and the sizes of recoverable systems. In any event, based on this situation, ACDC is not applicable to all sizes of software systems.
\subsubsection{ARC}
What we said in section \ref{ACDC_feasibility} also applies to ARC. We have observed that ARC exits attempts to recover the current version of Android with a java.lang.OutOfMemoryError exception, indicating that its virtual machine has run out of memory. Similar to ACDC, this seems to be caused by the size of Android, since we have not made any similar observations on smaller systems.
\subsubsection{Summary}
While PKG is feasible to run on any system, the results for ACDC and ARC are variable and unpredictable without a guarantee that a result will be produced at all.
\subsection{RQ3 (Clarity)}
For clarity, we looked at whether a user receives an explanation for why a given entity ended up in a specific cluster, unless it is immediately obvious.
\subsubsection{PKG}
Since for this recovery method, all code entities are grouped and clustered by packages, no further explanation is needed.
\subsubsection{ACDC}
No explanation is provided that would explain which specific programming pattern or set of patterns determined the grouping of a given entity into a specific cluster. Since this is not obvious, the user will not have this information.
\subsubsection{ARC}
ARC does not output its generated topics in one or more files, but they are displayed on the standard output while running it and can be captured from there.
Furthermore, it does not explain which topic an entity was clustered by, only providing the number of the topic. (It would be possible for the user to relate the numbers to the topics displayed in the standard output, but this functionality would have to be provided by the user) Even if it did so, in order for the reasons of the clustering to be clear the topics would have to be understandable, which is not always the case, as we can see in Section \ref{arc_eval}.
\subsection{RQ4 (Proportionality)}
Similarly to our considerations on feasibility (see section \ref{feasibility}), instead of committing to what exactly would be required of a result in order to be considered appropriately proportional to the changes in the input, we consider situations in which this would not be the case. One such situation is if the same, unchanged input can lead to significantly different results. Note that this requirement is stronger than nondeterminism alone, which may be locally contained or only lead to a small difference when measured by the methods described in Section \ref{sim_measures}.
Another case where a change in the result is disproportionate is when the results are deterministic, but labile, i.e., they change profoundly when only one bit is changed in the input.
\subsubsection{PKG}
PKG exhibits no disproportionate change in its output for changed input. 
\subsubsection{ACDC}
Similar to PKG, ACDC did not react disproportionately to minor changes in the input.
\subsubsection{ARC}\label{proportionality_arc}
In order to check the proportionality of ARC, we changed one letter in a comment in a arbitrary source file. The MojoFM value between the recovery of the original system and that of the changed system was 61.4\%, corresponding to a change of 38.6\%. The value for a2a was close to 90\%, the change being 10\%. This is out of the expected proportion for changing a single character.
\subsection{RQ5 (Determinism)}
\subsubsection{PKG}
PKG is deterministic for the reasons laid out in Section \ref{feasibility_pkg}.
\subsubsection{ACDC}
ACDC as implemented in ARCADE is deterministic.
\subsubsection{ARC}
ARC as implemented in ARC should be deterministic considering the fixes made to it, but surprisingly, our empirical results have shown it to be non-deterministic. MojoFM values between the same version of Apache Chukwa have had an average value of less than 72\%, meaning that they differ by more than 28\%. The average for a2a is 89.3\%, indicating a difference of 10.7\%.
\subsection{RQ6 (Continuity)} 
For continuity, we evaluate whether it is possible to go from one state to another through a series of intermediate states that show a ``direction''. In other words, if we follow the paradigm of the recovery methods, is it possible to do partial work that is reflected in the resulting architectural view?
\subsubsection{PKG}
Continuity is possible. Consider moving two entities from package A to package B, one at a time. In the initial state, both entities would be clustered under package A. In the final state, they would both be under package B's cluster. In the incomplete state when only the first component is moved, that component will be clustered in B, while the second one remains in A.
\subsubsection{ACDC}
While the lack of clarity of ACDC results makes this uncertain for all cases, continuity should be possible for small modifications.
\subsubsection{ARC}
This is not possible for the same reasons that apply to ACDC.
\subsection{RQ7 (Isolation, Additivity, Predictability)}
Here we look for \textit{crosstalk}, which means that changes in one part of the systems can have an effect on an unrelated one. 
\subsubsection{PKG}
Adding, moving or deleting entities only affects the packages in which it happens, therefore there is no crosstalk.
\subsubsection{ACDC}
The exact impact of a change cannot always be predicted, given that ACDC does not explain its results and which combination of patterns most influenced its result.
\subsubsection{ARC}
As we have already seen further above (see Section \ref{proportionality_arc}, any change in the system, no matter how small, leads to a considerably different view that reorganizes clusters everywhere. It is therefore not possible to isolate change in ARC.
\subsection{Topic Quality}
\subsubsection{Evaluated Systems} \label{arc_eval}
We have evaluated versions of Apache Hadoop, Android and Apache Chukwa.

Apache Hadoop 3.1.1 is the current version of Apache Hadoop as of this writing.

Since space considerations make showing all top words in the 100 topics generated for it prohibitve, the following is a relevant sample:

\begin{enumerate}
\item qe uu vv nb eq xx unie ip ey wf dd wr wo wn ww kw mb kg sn uy oz cv cg vy yr db jg ky ov ty aa um te wy ed ib ob ae nn xr de zm sy mm io wu le ue vm
\item xe xa xb xd xf xc xbb xdb xcb xee xbf xad xff xef xba xfd xfe xdc xeb xcd xea xaf xfb xcc xec xaa xca xbc xed xfa xcf xdf xfc xdd xbd xce xac xbe xda xab xae xde ca fe bt cf fu da fs
\item version resolved https registry tgz yarnpkg dependencies bb cc df cb cd ab ff dc de da dd ad fe fc fb ae aa ed ef bf fd ea db eb ce bd ba af bc fa cf ee ac crc ec ca
\item file files set data configuration directory default user cluster number support list change note html system time
\item path file fs dir filesystem directory delete files filestatus exists create
\item path uri fs filesystem final fc ioexception throws hadoop file src filestatus filecontext
\item path json response class string mediatype exception application jsonobject
\item public return override string
\item public return override private void import long boolean null int class super protected string apache extends
\item hadoop code apache org file
\item hadoop code file apache org path fs
\item hadoop org apache jar
\item hadoop jira apache issues https org
\item hadoop dir home classpath
\end{enumerate}

Immediate observations include that
\begin{itemize}
\item The first three topics contain combinations of two or three letters that will be meaningless at least to outsiders and newcomers to Hadoop.
\item The next three topics appear to be related to file handling.
\item The presence of ``uri'' in the sixth and ``json'', ``response'' and ``jsonobject'' in the seventh object make seem likely that ``path'' may be used in different context in different topics, considering that it can mean a path to a file in a filesystem or a path to a resource on the web. (Note that all words in all topics are converted to lowercase.)
\item The most prominent words in topics 8 and 9 are Java keywords.
\item The remaining five topics all start with ``hadoop''. (This word appears in 67 of 100 topics overall.)
\end{itemize}

The following is a selection of 5 from 30 concerns that were automatically recovered by ARC when recovering Apache Chukwa 0.7.0\cite{TheApacheSoftwareFoundation2016}:

\begin{itemize}
\item apach file org softwar hadoop chukwa write http addit inform copi notic foundat basi agre contributor permiss specif complianc express 
\item file path day chukwa hour apach record hadoop sourc org format cluster log conf merg dest status roll folder job 
\item file chunk chukwa apach test org hadoop conf writer key archiv impl configur stream cluster sequenc print reader seq util
\item path conf log file status error org apach chukwa configur loader hadoop param util uri logger nagio level factori warn
\item file path dir log conf chukwa read local delet directori apach hdfs gold len configur write hadoop absolut exist sys
\end{itemize}

\begin{figure}[t]
\centering
\includegraphics[width=0.39\textwidth]{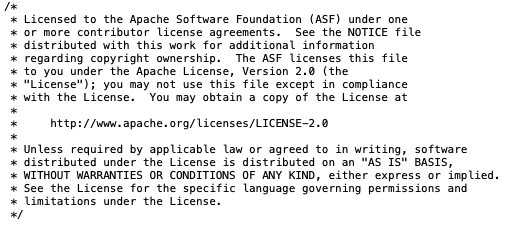}
\caption{The Apache License, Version 2.0}
\label{fig_apache_license}
\end{figure}

Immediate observations include that
\begin{enumerate}
\item All five topics shown above prominently include the words ``file'', ``apach'', ``chukwa'' and ``hadoop''.
\item All five topics seem to related to writing or loggin files. 
\item 17 topics contain the word ``chukwa'', 13 contain ``hadoop''. (Space considerations prohibit showing all topics here.)  
\item Many words in the first topic correspond to words or parts of words from the apache software license that Chukwa is distributed under (see Figure \ref{fig_apache_license}). These occurences include ``softwar'', ``inform'', ``notic'', ``foundat'', ``agre'', ``contributor'', ``permiss'', ``compliance'' and ``express''.
\end{enumerate}

\subsubsection{Results applying to all systems}
All systems exhibit common phenomena. 
\begin{enumerate}
\item In cases where several or even all source code entities  feature software license notices, one or more topics will be formed from it. (This is due to these words occurring together frequently.)
\item Some topics cannot be assigned a meaning to (such as the first three topics for Apache Hadoop).
\item The same concern, such as file handling, can be split over several topics.
\item The name of the system is prominently featured in several concerns.
\item Topics are formed from words that are incidental to a system and its functionality, such as licensing or the name of the system. Such topics cannot be regarded as a concern since they are not related to the system's role, responsibility, concept, or purpose (compare to the definition of a concern Section \ref{foundation}).
\end{enumerate}

The basic premise of ARC is that each detected topic maps to one concern. As has been shown above, this premise does not hold. Due to the basic building blocks being flawed, it is not possible to use an automatic run of ARC as the basis for an architectural view that reflects the concerns present in a system. Further conclusions from the architectural view are just as unwarranted.
 \section{Threats to Validity}\label{threats}
\subsection{Selection of recovery methods}
Our selection of recovery methods has been limited mostly due to practicalities such as the availability of working implementations. However, one motivation of our research was to investigate how suitable recovery methods that have been highly lauded might be for maintenance. This reduces the interest in recovery methods that are either less acclaimed or not evaluated in comparisons such as in \cite{Garcia2013}.
\subsection{Selection of recovered systems}
Any selection of systems is limited and may not be representative. However, since our reasoning about the attributes of each recovery method was deduced from the foundations of how these recovery methods work, we believe it should apply to any system and that our empirical results are only reality checks.
\subsection{Number of topics selected for ARC}

Depending on the respective systems and versions evaluated, setting the amount of topics to be recovered to 100 may be regarded as arbitrary high or low and therefor unfavorable. However, ARC does not contain any guidance on how many topics to start out with for a given system. After a given run, indications may be taken that the number of topics is too large or too small, and it may be iteratively adjusted over subsequent runs, but the literature on ARC does not mandate such a process, and we have not seen any indication that it was followed by studies that based conclusions on its results.
For this reason, our conclusions about its shortcomings remain. \section{Conclusion}\label{conclusion}
\subsection{ACDC}

How useful ACDC is for the maintenance of a given system depends on whether circumstances that the user cannot predict are beneficent. Even whether or not its results are architectural in nature can vary from one case to another. This not only forces the users to make this determination, but also makes it impossible to rely on using automatically gained architectural views from it as a basis for further processing.

\subsection{ARC}

Several issues exist that impact the quality of recovery results of ARC:
\begin{itemize}
\item Low-quality topics with no discernible meaning
\item Code entities are assigned extraneous concerns such as license wording (which is not addressed though programming)
\item Repetitive topics
\item Topics formed from words like the name of the system (which should be a stop word)
\end{itemize}
The deficiencies of the topics render clusters or architectural smells determined based on them questionable.
Consider the examples of the architectural smells Scattered Parasitic Functionality \cite{Garcia2009} and Concern Overload \cite{Le2016} Scattered Parasitic Functionality occurs when multiple components realize the same high-level concern while at least one of them also realizes an orthogonal concern. Detecting such a situation requires that the shared concern as well as the potentially orthogonal one are meaningful. As we have shown, this can not only not be guaranteed, but is seldom the case. Even if this would pose no problem, in order to attest this smell, the second concern would have to be orthogonal to the first one. Leaving out the question of how this could be automatically determined for two arbitrarily generated topics, it is clear that since a topic cannot be orthogonal to itself, two topics that have the same meaning can not be orthogonal to each other. 

As shown in Sections \ref{foundation} and \ref{results}, there is no way to guarantee that topic modeling will automatically come up with topics that are meaningful and different from each other.
The same issues also affect the detection of Concern Overload, which occurs when a component serves too many different concerns, i.e. the number of different concerns rises above a threshold. An overload of a component would require that the concerns are guaranteed to be actually different from each other, since otherwise they may be reduced to a number below the threshold used for this smell or even to one.
While it is possible to mitigate some of the deficiencies of the topics of ARC by closely supervising its topic generations, this would add an unpredictable about of time to its workflow.

\subsection{PKG}\label{conclusion_pkg}
PKG fulfills all our criteria for a value aid in maintenance, save for the one which requires its result to represent an architecture. This makes it unsuitable as an architecture recovery method, but leaves it usable to explore structural differences between different system versions. Its plethora of desirable qualities make it an interesting candidate for expansion into a full-fledged architecture recovery method.

\subsection{Summary}
All three recovery methods bring either potentially interesting views or desirable maintenance attributes to the table, but none of them has everything. The accuracy of the architectural views from ACDC and ARC remains hard to assess.
None of the three recovery methods allows maintainers to establish the virtuous cycle shown in Figure \ref{fig:virtuouscycle}. \section{Future Work}\label{future}
Since none of the recovery methods we considered fulfilled all or most of the criteria outlined in the introduction while also producing a meaningful architecture, our goal is to develop such a recovery method. A possible starting point for such a recovery method could be an extension of PKG, that makes it produce a more meaningful recovery result than just a package structure while also keeping its considerable number of desirable properties as outlined in Section \ref{conclusion_pkg} intact. One approach under consideration is to enrich the results of PKG with elements from ACDC and ARC, such as programming patterns and concerns.

The observed situations in which even comparably high memory resources were not enough to recover the popular Android system, it is also of interest to us to develop a recovery method that can build up its recovery results from smaller parts that are computed in a distributed manner and then combine the partial results.  
\bibliographystyle{ACM-Reference-Format}

\begin{thebibliography}{35}



\ifx \showCODEN    \undefined \def \showCODEN     #1{\unskip}     \fi
\ifx \showDOI      \undefined \def \showDOI       #1{#1}\fi
\ifx \showISBNx    \undefined \def \showISBNx     #1{\unskip}     \fi
\ifx \showISBNxiii \undefined \def \showISBNxiii  #1{\unskip}     \fi
\ifx \showISSN     \undefined \def \showISSN      #1{\unskip}     \fi
\ifx \showLCCN     \undefined \def \showLCCN      #1{\unskip}     \fi
\ifx \shownote     \undefined \def \shownote      #1{#1}          \fi
\ifx \showarticletitle \undefined \def \showarticletitle #1{#1}   \fi
\ifx \showURL      \undefined \def \showURL       {\relax}        \fi
\providecommand\bibfield[2]{#2}
\providecommand\bibinfo[2]{#2}
\providecommand\natexlab[1]{#1}
\providecommand\showeprint[2][]{arXiv:#2}

\bibitem[\protect\citeauthoryear{Andritsos and Tzerpos}{Andritsos and
  Tzerpos}{2005}]        {Andritsos2005}
\bibfield{author}{\bibinfo{person}{Periklis Andritsos} {and}
  \bibinfo{person}{Vassilios Tzerpos}.} \bibinfo{year}{2005}\natexlab{}.
\newblock \showarticletitle{{Information-Theoretic Software Clustering}}.
\newblock \bibinfo{journal}{\emph{IEEE Transactions on Software Engineering}}
  \bibinfo{volume}{31}, \bibinfo{number}{2} (\bibinfo{year}{2005}),
  \bibinfo{pages}{150--165}.
\newblock


\bibitem[\protect\citeauthoryear{Apache}{Apache}{2018}]        {Apache2018}
\bibfield{author}{\bibinfo{person}{Apache}.} \bibinfo{year}{2018}\natexlab{}.
\newblock \bibinfo{title}{{Apache Hadoop}}.
\newblock
\newblock
\urldef\tempurl\url{http://hadoop.apache.org/}
\showURL{\tempurl}


\bibitem[\protect\citeauthoryear{Behnamghader, Le, Garcia, Link, Shahbazian,
  and Medvidovic}{Behnamghader et~al\mbox{.}}{2017}]        {Behnamghader2016}
\bibfield{author}{\bibinfo{person}{Pooyan Behnamghader},
  \bibinfo{person}{Duc~Minh Le}, \bibinfo{person}{Joshua Garcia},
  \bibinfo{person}{Daniel Link}, \bibinfo{person}{Arman Shahbazian}, {and}
  \bibinfo{person}{Nenad Medvidovic}.} \bibinfo{year}{2017}\natexlab{}.
\newblock \showarticletitle{A large-scale study of architectural evolution in
  open-source software systems}.
\newblock \bibinfo{journal}{\emph{Empirical Software Engineering}}
  \bibinfo{volume}{22}, \bibinfo{number}{3} (\bibinfo{year}{2017}),
  \bibinfo{pages}{1146--1193}.
\newblock


\bibitem[\protect\citeauthoryear{Blei}{Blei}{2012}]        {Blei2012}
\bibfield{author}{\bibinfo{person}{David~M. Blei}.}
  \bibinfo{year}{2012}\natexlab{}.
\newblock \showarticletitle{{Introduction to Probabilistic Topic Modeling}}.
\newblock \bibinfo{journal}{\emph{Commun. ACM}}  \bibinfo{volume}{55}
  (\bibinfo{year}{2012}), \bibinfo{pages}{77--84}.
\newblock
\showISBNx{0001-0782}
\showISSN{00010782}
\urldef\tempurl\url{https://doi.org/10.1145/2133806.2133826}
\showDOI{\tempurl}
\showeprint[arxiv]{1003.4916}


\bibitem[\protect\citeauthoryear{Blei, Ng, and Jordan}{Blei
  et~al\mbox{.}}{2003}]        {Blei2003}
\bibfield{author}{\bibinfo{person}{David~M. Blei}, \bibinfo{person}{Andrew
  Y.~A.Y. Ng}, {and} \bibinfo{person}{M.I. Michael~I. Jordan}.}
  \bibinfo{year}{2003}\natexlab{}.
\newblock \showarticletitle{{Latent dirichlet allocation}}.
\newblock \bibinfo{journal}{\emph{Neural Information Processing Systems}}
  \bibinfo{volume}{3}, \bibinfo{number}{4-5} (\bibinfo{year}{2003}),
  \bibinfo{pages}{993--1022}.
\newblock
\showISBNx{9781577352815}
\showISSN{1532-4435}
\urldef\tempurl\url{https://doi.org/10.1162/jmlr.2003.3.4-5.993}
\showDOI{\tempurl}
\showeprint[arxiv]{1111.6189v1}


\bibitem[\protect\citeauthoryear{Bowman, Holt, and Brewster}{Bowman
  et~al\mbox{.}}{1999}]        {Bowman1999}
\bibfield{author}{\bibinfo{person}{Ivan~T Bowman}, \bibinfo{person}{Richard~C
  Holt}, {and} \bibinfo{person}{Neil~V Brewster}.}
  \bibinfo{year}{1999}\natexlab{}.
\newblock \showarticletitle{Linux as a case study: Its extracted software
  architecture}.
\newblock  (\bibinfo{year}{1999}), \bibinfo{pages}{555--563}.
\newblock


\bibitem[\protect\citeauthoryear{{Computer Science Department}}{{Computer
  Science Department}}{2017}]        {ComputerScienceDepartment2017}
\bibfield{author}{\bibinfo{person}{University of Southern~California {Computer
  Science Department}}.} \bibinfo{year}{2017}\natexlab{}.
\newblock \bibinfo{booktitle}{\emph{{ARCADE Manual}}}.
\newblock \bibinfo{publisher}{University of Southern California},
  \bibinfo{address}{Los Angeles}.
\newblock
\urldef\tempurl\url{https://softarch.usc.edu/~lemduc/Recovered_files/ArchitectureEvolutionAnalysiswithARCADE.pdf}
\showURL{\tempurl}


\bibitem[\protect\citeauthoryear{Elmer}{Elmer}{2014}]        {Elmer2014}
\bibfield{author}{\bibinfo{person}{Franz-Josef Elmer}.}
  \bibinfo{year}{2014}\natexlab{}.
\newblock \bibinfo{title}{{Classycle: Analysing Tools for Java Class and
  Package Dependencies}}.
\newblock
\newblock
\urldef\tempurl\url{http://classycle.sourceforge.net}
\showURL{\tempurl}


\bibitem[\protect\citeauthoryear{Emden}{Emden}{2002}]        {Emden2002}
\bibfield{author}{\bibinfo{person}{Eva~Van Emden}.}
  \bibinfo{year}{2002}\natexlab{}.
\newblock \showarticletitle{{Software Quality Assurance by Detecting Code
  Smells}}.
\newblock \bibinfo{journal}{\emph{Informatica}} (\bibinfo{year}{2002}),
  \bibinfo{pages}{16--17}.
\newblock


\bibitem[\protect\citeauthoryear{Fakhroutdinov}{Fakhroutdinov}{2013}]        {fakhroutdinov_2013}
\bibfield{author}{\bibinfo{person}{Kirill Fakhroutdinov}.}
  \bibinfo{year}{2013}\natexlab{}.
\newblock \bibinfo{title}{{Web Application Clusters UML Deployment Diagram
  Example}}.
\newblock
\newblock
\urldef\tempurl\url{https://www.uml-diagrams.org//web-application-clusters-uml-deployment-diagram-example.html}
\showURL{\tempurl}


\bibitem[\protect\citeauthoryear{Garcia}{Garcia}{2014}]        {Garcia2014}
\bibfield{author}{\bibinfo{person}{Joshua Garcia}.}
  \bibinfo{year}{2014}\natexlab{}.
\newblock \emph{\bibinfo{title}{{A Unified Framework for Studying Architectural
  Decay of Software Systems}}}.
\newblock Dissertation. \bibinfo{school}{University of Southern California}.
\newblock


\bibitem[\protect\citeauthoryear{Garcia}{Garcia}{2018}]        {Garcia2018}
\bibfield{author}{\bibinfo{person}{Joshua Garcia}.}
  \bibinfo{year}{2018}\natexlab{}.
\newblock \bibinfo{title}{{Arcade}}.
\newblock
\newblock
\urldef\tempurl\url{https://bitbucket.org/joshuaga/arcade/src/master/}
\showURL{\tempurl}


\bibitem[\protect\citeauthoryear{Garcia, Ivkovic, and Medvidovic}{Garcia
  et~al\mbox{.}}{2013}]        {Garcia2013}
\bibfield{author}{\bibinfo{person}{Joshua Garcia}, \bibinfo{person}{Igor
  Ivkovic}, {and} \bibinfo{person}{Nenad Medvidovic}.}
  \bibinfo{year}{2013}\natexlab{}.
\newblock \showarticletitle{A comparative analysis of software architecture
  recovery techniques}. In \bibinfo{booktitle}{\emph{Proceedings of the 28th
  IEEE/ACM International Conference on Automated Software Engineering}}. IEEE
  Press, \bibinfo{pages}{486--496}.
\newblock


\bibitem[\protect\citeauthoryear{Garcia, Popescu, Edwards, and
  Medvidovic}{Garcia et~al\mbox{.}}{2009}]        {Garcia2009}
\bibfield{author}{\bibinfo{person}{J Garcia}, \bibinfo{person}{D Popescu},
  \bibinfo{person}{G Edwards}, {and} \bibinfo{person}{N Medvidovic}.}
  \bibinfo{year}{2009}\natexlab{}.
\newblock \showarticletitle{{Identifying Architectural Bad Smells}}.
\newblock \bibinfo{journal}{\emph{2009 13th European Conference on Software
  Maintenance and Reengineering}} (\bibinfo{year}{2009}),
  \bibinfo{pages}{255--258}.
\newblock
\showISBNx{978-1-4244-3755-9}
\showISSN{1534-5351}
\urldef\tempurl\url{https://doi.org/10.1109/CSMR.2009.59}
\showDOI{\tempurl}


\bibitem[\protect\citeauthoryear{Garcia, Popescu, Mattmann, Medvidovic, and
  Cai}{Garcia et~al\mbox{.}}{2011}]        {Garcia2011}
\bibfield{author}{\bibinfo{person}{Joshua Garcia}, \bibinfo{person}{Daniel
  Popescu}, \bibinfo{person}{Chris Mattmann}, \bibinfo{person}{Nenad
  Medvidovic}, {and} \bibinfo{person}{Yuanfang Cai}.}
  \bibinfo{year}{2011}\natexlab{}.
\newblock \showarticletitle{{Enhancing architectural recovery using concerns}}.
  In \bibinfo{booktitle}{\emph{2011 26th IEEE/ACM International Conference on
  Automated Software Engineering, ASE 2011, Proceedings}}.
  \bibinfo{pages}{552--555}.
\newblock
\showISBNx{9781457716393}
\showISSN{1938-4300}
\urldef\tempurl\url{https://doi.org/10.1109/ASE.2011.6100123}
\showDOI{\tempurl}


\bibitem[\protect\citeauthoryear{Google}{Google}{2018}]        {Google2018}
\bibfield{author}{\bibinfo{person}{Google}.} \bibinfo{year}{2018}\natexlab{}.
\newblock \bibinfo{title}{{Android Open Source Project}}.
\newblock
\newblock
\urldef\tempurl\url{https://source.android.com}
\showURL{\tempurl}


\bibitem[\protect\citeauthoryear{Herbsleb and Mockus}{Herbsleb and
  Mockus}{2003}]        {Herbsleb2003}
\bibfield{author}{\bibinfo{person}{James~D Herbsleb} {and}
  \bibinfo{person}{Audris Mockus}.} \bibinfo{year}{2003}\natexlab{}.
\newblock \showarticletitle{{An Empirical Study of Speed and Communication in
  Globally Distributed Software Development}}.
\newblock \bibinfo{journal}{\emph{IEEE Transactions on Software Engineering}}
  \bibinfo{volume}{29}, \bibinfo{number}{6} (\bibinfo{year}{2003}),
  \bibinfo{pages}{481--494}.
\newblock
\urldef\tempurl\url{https://doi.org/10.1109/TSE.2003.1205177}
\showDOI{\tempurl}


\bibitem[\protect\citeauthoryear{ISO/IEC/IEEE}{ISO/IEC/IEEE}{2011}]        {ISO/IEC/IEEE}
\bibfield{author}{\bibinfo{person}{ISO/IEC/IEEE}.}
  \bibinfo{year}{2011}\natexlab{}.
\newblock \bibinfo{title}{{Systems and software engineering-- Architecture
  description}}.
\newblock , \bibinfo{numpages}{37}~pages.
\newblock
\urldef\tempurl\url{http://www.iso-architecture.org/42010/defining-architecture.html}
\showURL{\tempurl}


\bibitem[\protect\citeauthoryear{Le, Behnamghader, Garcia, Link, Shahbazian,
  and Medvidovic}{Le et~al\mbox{.}}{2015}]        {Le2015}
\bibfield{author}{\bibinfo{person}{Duc~Minh Le}, \bibinfo{person}{Pooyan
  Behnamghader}, \bibinfo{person}{Joshua Garcia}, \bibinfo{person}{Daniel
  Link}, \bibinfo{person}{Arman Shahbazian}, {and} \bibinfo{person}{Nenad
  Medvidovic}.} \bibinfo{year}{2015}\natexlab{}.
\newblock \showarticletitle{{An empirical study of architectural change in
  open-source software systems}}.
\newblock \bibinfo{journal}{\emph{IEEE International Working Conference on
  Mining Software Repositories}}  \bibinfo{volume}{2015-Augus}
  (\bibinfo{year}{2015}), \bibinfo{pages}{235--245}.
\newblock
\showISBNx{9780769555942}
\showISSN{21601860}
\urldef\tempurl\url{https://doi.org/10.1109/MSR.2015.29}
\showDOI{\tempurl}


\bibitem[\protect\citeauthoryear{Le, Carrillo, Capilla, and Medvidovic}{Le
  et~al\mbox{.}}{2016}]        {Le2016}
\bibfield{author}{\bibinfo{person}{Duc~Minh Le}, \bibinfo{person}{Carlos
  Carrillo}, \bibinfo{person}{Rafael Capilla}, {and} \bibinfo{person}{Nenad
  Medvidovic}.} \bibinfo{year}{2016}\natexlab{}.
\newblock \showarticletitle{Relating architectural decay and sustainability of
  software systems}. In \bibinfo{booktitle}{\emph{Software Architecture
  (WICSA), 2016 13th Working IEEE/IFIP Conference on}}. IEEE,
  \bibinfo{pages}{178--181}.
\newblock


\bibitem[\protect\citeauthoryear{Le, Link, Shahbazian, and Medvidovic}{Le
  et~al\mbox{.}}{2018}]        {Le2018}
\bibfield{author}{\bibinfo{person}{Duc~Minh Le}, \bibinfo{person}{Daniel Link},
  \bibinfo{person}{Arman Shahbazian}, {and} \bibinfo{person}{Nenad
  Medvidovic}.} \bibinfo{year}{2018}\natexlab{}.
\newblock \showarticletitle{{An Empirical Study of Architectural Decay in
  Open-Source Software}}.
\newblock \bibinfo{journal}{\emph{Proceedings - 2018 IEEE 15th International
  Conference on Software Architecture, ICSA 2018}} (\bibinfo{year}{2018}),
  \bibinfo{pages}{176--185}.
\newblock
\showISBNx{9781538663981}
\urldef\tempurl\url{https://doi.org/10.1109/ICSA.2018.00027}
\showDOI{\tempurl}


\bibitem[\protect\citeauthoryear{Lutellier, Chollak, Garcia, Tan, Rayside,
  Medvidovic, and Kroeger}{Lutellier et~al\mbox{.}}{2015}]        {Lutellier2015}
\bibfield{author}{\bibinfo{person}{Thibaud Lutellier}, \bibinfo{person}{Devin
  Chollak}, \bibinfo{person}{Joshua Garcia}, \bibinfo{person}{Lin Tan},
  \bibinfo{person}{Derek Rayside}, \bibinfo{person}{Nenad Medvidovic}, {and}
  \bibinfo{person}{Robert Kroeger}.} \bibinfo{year}{2015}\natexlab{}.
\newblock \showarticletitle{{Comparing Software Architecture Recovery
  Techniques Using Accurate Dependencies}}.
\newblock \bibinfo{journal}{\emph{Proceedings - International Conference on
  Software Engineering}}  \bibinfo{volume}{2} (\bibinfo{year}{2015}),
  \bibinfo{pages}{69--78}.
\newblock
\showISBNx{9781479919345}
\showISSN{02705257}
\urldef\tempurl\url{https://doi.org/10.1109/ICSE.2015.136}
\showDOI{\tempurl}


\bibitem[\protect\citeauthoryear{Makrehchi and Kamel}{Makrehchi and
  Kamel}{2008}]        {Makrehchi2008}
\bibfield{author}{\bibinfo{person}{Masoud Makrehchi} {and}
  \bibinfo{person}{Mohamed~S Kamel}.} \bibinfo{year}{2008}\natexlab{}.
\newblock \showarticletitle{Automatic extraction of domain-specific stopwords
  from labeled documents}.
\newblock  (\bibinfo{year}{2008}), \bibinfo{pages}{222--233}.
\newblock


\bibitem[\protect\citeauthoryear{Mancoridis, Mitchell, Chen, and
  Gansner}{Mancoridis et~al\mbox{.}}{1999}]        {Mancoridis1999}
\bibfield{author}{\bibinfo{person}{S. Mancoridis}, \bibinfo{person}{B.S.
  Mitchell}, \bibinfo{person}{Y. Chen}, {and} \bibinfo{person}{E.R. Gansner}.}
  \bibinfo{year}{1999}\natexlab{}.
\newblock \showarticletitle{{Bunch: a clustering tool for the recovery and
  maintenance of software system structures}}.
\newblock \bibinfo{journal}{\emph{Proceedings IEEE International Conference on
  Software Maintenance - 1999 (ICSM'99). 'Software Maintenance for Business
  Change' (Cat. No.99CB36360)}} (\bibinfo{year}{1999}),
  \bibinfo{pages}{50--59}.
\newblock
\showISBNx{0-7695-0016-1}
\showISSN{1063-6773}
\urldef\tempurl\url{https://doi.org/10.1109/ICSM.1999.792498}
\showDOI{\tempurl}


\bibitem[\protect\citeauthoryear{Mattmann, Garcia, Krka, Popescu, and
  Medvidovic}{Mattmann et~al\mbox{.}}{2015}]        {Mattmann2015a}
\bibfield{author}{\bibinfo{person}{Chris~A. Mattmann}, \bibinfo{person}{Joshua
  Garcia}, \bibinfo{person}{Ivo Krka}, \bibinfo{person}{Daniel Popescu}, {and}
  \bibinfo{person}{Nenad Medvidovic}.} \bibinfo{year}{2015}\natexlab{}.
\newblock \showarticletitle{{Revisiting the Anatomy and Physiology of the
  Grid}}.
\newblock \bibinfo{journal}{\emph{Journal of Grid Computing}}
  \bibinfo{volume}{13}, \bibinfo{number}{1} (\bibinfo{year}{2015}),
  \bibinfo{pages}{19--34}.
\newblock
\showISBNx{9781424449859}
\showISSN{15707873}
\urldef\tempurl\url{https://doi.org/10.1007/s10723-015-9324-0}
\showDOI{\tempurl}


\bibitem[\protect\citeauthoryear{Merriam-Webster}{Merriam-Webster}{2018}]        {Merriam-Webster}
\bibfield{author}{\bibinfo{person}{Merriam-Webster}.}
  \bibinfo{year}{2018}\natexlab{}.
\newblock \bibinfo{title}{{Concern | Definition of Concern by
  Merriam-Webster}}.
\newblock
\newblock
\urldef\tempurl\url{https://www.merriam-webster.com/dictionary/concern}
\showURL{\tempurl}


\bibitem[\protect\citeauthoryear{{Mockus, Audris and T Fielding, Roy and D
  Herbsleb}}{{Mockus, Audris and T Fielding, Roy and D Herbsleb}}{2002}]        {MockusAudrisandTFieldingRoyandDHerbsleb2002}
\bibfield{author}{\bibinfo{person}{James {Mockus, Audris and T Fielding, Roy
  and D Herbsleb}}.} \bibinfo{year}{2002}\natexlab{}.
\newblock \showarticletitle{{Two Case Studies of Open Source Software
  Development: Apache and Mozilla}}.
\newblock \bibinfo{journal}{\emph{ACM Transactions on Software Engineering and
  Methodology}} \bibinfo{volume}{11}, \bibinfo{number}{3}
  (\bibinfo{year}{2002}), \bibinfo{pages}{309----346}.
\newblock


\bibitem[\protect\citeauthoryear{Mohagheghi, Anda, and Conradi}{Mohagheghi
  et~al\mbox{.}}{2005}]        {Mohagheghi2005}
\bibfield{author}{\bibinfo{person}{Parastoo Mohagheghi}, \bibinfo{person}{Bente
  Anda}, {and} \bibinfo{person}{Reidar Conradi}.}
  \bibinfo{year}{2005}\natexlab{}.
\newblock \showarticletitle{{Effort estimation of use cases for incremental
  large-scale software development}}.
\newblock \bibinfo{journal}{\emph{Proceedings of the 27th international
  conference on Software engineering - ICSE '05}} \bibinfo{number}{June}
  (\bibinfo{year}{2005}), \bibinfo{pages}{303}.
\newblock
\showISBNx{1595939632}
\showISSN{14779234}
\urldef\tempurl\url{https://doi.org/10.1145/1062455.1062516}
\showDOI{\tempurl}


\bibitem[\protect\citeauthoryear{Shahbazian, Lee, Le, Brun, and
  Medvidovic}{Shahbazian et~al\mbox{.}}{2018}]        {Shahbazian2017}
\bibfield{author}{\bibinfo{person}{Arman Shahbazian}, \bibinfo{person}{Youn~Kyu
  Lee}, \bibinfo{person}{Duc Le}, \bibinfo{person}{Yuriy Brun}, {and}
  \bibinfo{person}{Nenad Medvidovic}.} \bibinfo{year}{2018}\natexlab{}.
\newblock \showarticletitle{Uncovering architectural design decisions}. In
  \bibinfo{booktitle}{\emph{2018 IEEE International Conference on Software
  Architecture (ICSA)}}. IEEE, \bibinfo{pages}{95--9509}.
\newblock


\bibitem[\protect\citeauthoryear{Shtern and Tzerpos}{Shtern and
  Tzerpos}{2010}]        {Shtern2010}
\bibfield{author}{\bibinfo{person}{Mark Shtern} {and}
  \bibinfo{person}{Vassilios Tzerpos}.} \bibinfo{year}{2010}\natexlab{}.
\newblock \bibinfo{title}{{ACDC Algorithm (Software Clustering)}}.
\newblock
\newblock
\urldef\tempurl\url{https://wiki.eecs.yorku.ca/project/cluster/protected:acdc}
\showURL{\tempurl}


\bibitem[\protect\citeauthoryear{Taylor, Medvidovic, and Dashofy}{Taylor
  et~al\mbox{.}}{2010}]        {Taylor2010}
\bibfield{author}{\bibinfo{person}{Richard~N. Taylor}, \bibinfo{person}{Nenad
  Medvidovic}, {and} \bibinfo{person}{Eric~M. Dashofy}.}
  \bibinfo{year}{2010}\natexlab{}.
\newblock \bibinfo{booktitle}{\emph{{Software Architecture - Foundations,
  Theory, and Practice}}}.
\newblock \bibinfo{publisher}{Wiley}.
\newblock


\bibitem[\protect\citeauthoryear{{The Apache Software Foundation}}{{The Apache
  Software Foundation}}{2016}]        {TheApacheSoftwareFoundation2016}
\bibfield{author}{\bibinfo{person}{{The Apache Software Foundation}}.}
  \bibinfo{year}{2016}\natexlab{}.
\newblock \bibinfo{title}{{Welcome To Apache Chukwa}}.
\newblock
\newblock
\urldef\tempurl\url{http://chukwa.apache.org}
\showURL{\tempurl}


\bibitem[\protect\citeauthoryear{{The Eclipse Foundation}}{{The Eclipse
  Foundation}}{2017}]        {TheEclipseFoundation2017}
\bibfield{author}{\bibinfo{person}{{The Eclipse Foundation}}.}
  \bibinfo{year}{2017}\natexlab{}.
\newblock \bibinfo{title}{{Eclipse}}.
\newblock
\newblock
\urldef\tempurl\url{https://www.eclipse.org}
\showURL{\tempurl}


\bibitem[\protect\citeauthoryear{Tzerpos and Holt}{Tzerpos and Holt}{2000}]        {tzerpos2000acdc}
\bibfield{author}{\bibinfo{person}{Vassilios Tzerpos} {and}
  \bibinfo{person}{Richard~C Holt}.} \bibinfo{year}{2000}\natexlab{}.
\newblock \showarticletitle{ACDC: An algorithm for comprehension-driven
  clustering}. In \bibinfo{booktitle}{\emph{Reverse Engineering, 2000.
  Proceedings. Seventh Working Conference on}}. IEEE,
  \bibinfo{pages}{258--267}.
\newblock


\bibitem[\protect\citeauthoryear{{Zhihua Wen} and Tzerpos}{{Zhihua Wen} and
  Tzerpos}{2004}]        {ZhihuaWen2004}
\bibfield{author}{\bibinfo{person}{{Zhihua Wen}} {and} \bibinfo{person}{V.
  Tzerpos}.} \bibinfo{year}{2004}\natexlab{}.
\newblock \showarticletitle{{An effectiveness measure for software clustering
  algorithms}}.
\newblock \bibinfo{journal}{\emph{Proceedings. 12th IEEE International Workshop
  on Program Comprehension, 2004.}} (\bibinfo{year}{2004}),
  \bibinfo{pages}{194--203}.
\newblock
\showISBNx{0-7695-2149-5}
\showISSN{1092-8138}
\urldef\tempurl\url{https://doi.org/10.1109/WPC.2004.1311061}
\showDOI{\tempurl}


\end{thebibliography}

\end{document}